\documentstyle[aps]{revtex}
\begin{document}
 \bibliographystyle{prsty}
\newcommand{\C}{\cite}
\newcommand{\beq}{\begin{equation}}
\newcommand{\eeq}{\end{equation}}
\newcommand{\bea}{\begin{eqnarray}}
\newcommand{\eea}{\end{eqnarray}}
\newcommand{\ms}{m_{\rm s}}
\newcommand{\MeV}{{\rm MeV}}
\newcommand{\fm}{{\rm fm}}
\newcommand{\alf}{{\bar a}}
\newcommand{\bet}{{\bar b}}
\newcommand{\lb}{\hfil\break }
\newcommand{\qeq}[1]{Eq.\ (\ref{#1})  }
\newcommand{\dsl}{ \rlap{/}{\partial} }
\newcommand{\pst}{ \rlap{/}{p}  }
\newcommand{\dm}{\Delta m}
\newcommand{\half}{\frac{1}{2}}
\newcommand{\quref}[1]{\cite{bolo:#1}}
\newcommand{\qref}[1]{Ref.\ \cite{bolo:#1}}
\newcommand{\queq}[1]{(\ref{#1})}
\newcommand{\qtab}[1]{Tab. \ref{#1}}
\newcommand{\wu}{\sqrt{3}}
\newcommand{\nn}{\nonumber \\ }
\newcommand{\Y}{\ {\cal Y}}
\newcommand{\Sp}{{\rm Sp\ } }
\newcommand{\Spto}{{\rm Sp_{(to)}\ } }
\newcommand{\Tr}{{\rm Tr\ } }
\newcommand{\tr}{{\rm tr\ } }
\newcommand{\sign}{{\rm sign} }
\newcommand{\linie}{\ \vrule height 14pt depth 7pt \ }
\newcommand{\intT}{\int_{-T/2}^{T/2} }
\newcommand{\Nc}{N_{\rm c}}
\newcommand{\Ne}{$N_{\rm c}$}
\newcommand{\gs}{$g_{\rm A}^{0}$}
\newcommand{\gt}{$g_{\rm A}^{3}$}
\newcommand{\go}{$g_{\rm A}^{8}$}
\newcommand{\Gs}{g_{\rm A}^{0}}
\newcommand{\Gt}{g_{\rm A}^{3}}
\newcommand{\Go}{g_{\rm A}^{8}}
\input epsf

\setcounter{page}{0}

\title
{\Large ~\\
{}~\\
Isospin Mass Splittings and the $\ms$ Corrections      \\
   in the  Semibosonized SU(3)-NJL-model}

\vspace{2 cm}

\author{
Andree Blotz$^{(1)}$
\footnote{email:andreeb@elektron.tp2.ruhr-uni-bochum.de},
K. Goeke$^{(1)}$
\footnote{email:goeke@hadron.tp2.ruhr-uni-bochum.de}
and
Micha{\l} Prasza{\l}owicz$^{(2)}$
\footnote{email:michal@thrisc.if.uj.edu.pl}}

\vspace{1cm}

\address{(1)
Institute for  Theoretical  Physics  II, \\  P.O. Box 102148,
Ruhr-University Bochum, \\
 D-W-44780 Bochum, Germany  \\
       }
\address{(2)
Institute of Physics,
Jagellonian University, \\  Reymonta 4, 30-059,   Krak{\'o}w, Poland
                         }
\date{ \today }
\maketitle

\vspace{3 cm}

\begin{abstract}
 The  mass splittings
of hyperons including the isospin splittings are calculated
with $O(\ms^2)$ and $O(\ms \dm)$ accuracy respectively
within the semibosonized SU(3)-NJL model. The pattern of the
isospin splittings is not spoiled by the terms of the order
$O(\ms \dm)$, and both splittings between the different isospin
multiplets and within the same multiplet are well reproduced
for acceptable values of $\ms$ and $\dm$.
\end{abstract}

\thispagestyle{empty}

\vspace{7 cm}
\begin{flushright}
RUB-TPII-6/94 \\
hep-ph/9409297
\end{flushright}

\newpage
\setcounter{page}{1}
\section{Introduction}

Although Quantum Chromodynamics (QCD), as the ultimate theory of strong
interactions, should be able to give precise predictions of all
physical quantities such as masses, mass splittings etc., its
predictive power is obscured by technical difficulties. One way
out is to employ effective models, which share some important
features with QCD. The chiral quark model \quref{dipepo} (or semibosonized
Nambu--Jona-Lasinio model) exhibits chiral symmetry breaking,
which, together with confinement, is relevant for the low energy
regime of QCD. Moreover it can be "derived" from QCD within the
instanton liquid model of the QCD vacuum \quref{dype1}.

The NJL model \quref{njl1} has
been therefore thoroughly examined against the low energy hadronic
data and it turned out to give surprisingly good predictions
for mass splittings \cite{bolo:ab4,bolo:wealre},  electromagnetic properties
\cite{bolo:allstars} and axial
properties of hyperons \cite{bolo:ab9,bolo:ab10}. The {\em hadronic} part of
the
isospin breaking effects which is  due to $\dm =m_{\rm d}-m_{\rm u}$
mass difference is  perhaps the most striking example of the
accuracy reached by the model. In \qref{ab5}  the isospin mass
differences have been calculated for the octet and decuplet of
baryons. Although the error bars on these splittings are large
(since the electromagnetic part of the splittings has to be
subtracted) the model predictions fall nicely within the error
bars for all spin 1/2 and 3/2 baryons for $\dm \approx 3.5$~MeV.

These results have been obtained within the perturbative approach,
where one expands the collective hamiltonian describing
baryonic states in terms of $\ms$ and $\dm$ up to terms linear in
 both parameters. It has been claimed in the literature
\C{bolo:Yabutalk,bolo:Stern}
that, at least within the Skyrme model approach, the next
 orders in $\ms$  spoil the nice pattern of the isospin
splittings obtained in the linear approximation. It is the purpose
of this work to examine if this is also  the case in the
present model. As will be
seen in the following, the inclusion of the terms of the order
$\dm\;\ms$ does not spoil the pattern of the isospin breaking,
however  $\dm$ which reproduces the experimental data is
shifted towards the higher value of the order of 4.5~MeV.
We consider this result as fully satisfactory. Other chiral
models either underestimate the isospin splittings by
factor of 2 or overestimate it by approximately the same
amount.\footnote{See Introduction in Ref.\C{bolo:Stern}}

The crucial point of our analysis is the fact that
though we are considering isospin breaking effects
of non-strange current masses we are using
the SU(3) version of the NJL model. In this framework the symmetry breaking
operator is proportional to $\lambda_3$. In the SU(2) model it
reduces to $\tau_3$ and  most of the matrix elements
after the collective quantization vanish. In SU(3) the polarization
of the strange Dirac sea provides a number of terms which are
crucial for the splitting pattern.

\section{Mesonic Sector}

The prominent feature of the NJL model consists in the fact that the
same effective action describes meson physics and also, through the
solitonic solutions, baryon physics. It is customary to fix the
parameters of the model by fitting meson properties. This procedure
leaves usually one free  parameter, namely the constituent quark mass $M$.
The regularization cutoff (or cutoff function) becomes then a function
of $M$. Then $M$ is fixed to fit {\em one} splitting in the soliton
sector, {\em eg.} N$-\Delta$, and then all other baryonic observables
come out as predictions. In practice there is always some freedom in
tuning other parameters like $\ms$ or $\dm$, although they are always
fairly constrained by the meson sector.

There are in fact different ways to treat the meson sector. One can
either solve the gap equations and then evaluate meson masses at zero
momentum $q=0$
from the curvature of the effective potential. One can also
evaluate the curvature at $q^2=-m_{\rm meson}^2$. Or
one can solve Bethe-Salpether equations for the meson propagators.
Another way is to fix
the parameters in the gradient expansion, which results in an effective
meson lagrangians whose coupling constants  are known from
the meson scattering. Each of these methods produces slightly different
results, which are not important for the gross features of the soliton
sector, but may influence the numerics. This influence is not very
significant if one is interested  in the quantities which are not too
small, however they might turn out important, if
one considers such tiny effects as the isospin splittings.

It is not our purpose to make a complete calculation of the isospin
splittings. That would require to include the electromagnetic
effects, which is certainly beyond the scope of this paper. Our goal is
by far more modest: we want to investigate the effect of the strange
quark mass on the hadronic part of the isospin splittings, very much in
the spirit of the calculations performed in the Skyrme model \quref{jjmps}.
Electromagnetic contributions will be simply parametrized by means
of some convenient Ansatz; in the meson sector we use the Dashen
Ansatz:
\beq
m_{\rm meson}^2 = m_{\rm H}^2 + {\cal Q}^2 c ,
\eeq
where ${\cal Q}$ is the meson charge, $c$ is a constant which
does not depend on the meson in question and subscript H stands
for the hadronic (quark mass dependent) part of the meson mass.

We fix the model parameters in the meson sector in the first
order in the quark mass matrix. In this way we avoid unnecessary
complications due to the $\eta^0-\eta^8-\pi^3$ mixing. We also
refrain from such problems as boson loops and the validity of the
Dashen Ansatz. All
these complications result in the uncertainty in the value of
$\dm$ extracted from the mesonic data.

The first order meson mass formulae with the Dashen Ansatz
have been analyzed in \qref{ab5}
They imply:
\beq
{\cal R}=\frac{\dm}{m_{\rm u}+m_{\rm d}}= 0.28.
\eeq
 For $m_{\rm u}+m_{\rm d}=12.2$ MeV, which is required by the
regularization prescription we use and which falls within the commonly
accepted range of values for $m_{\rm u}+m_{\rm d}$, we get $\dm=3.4$ MeV.
Ignoring completely electromagnetic contributions
would yield ${\cal R}=0.21$ and   $\dm=2.6$ MeV.

It should be
however stressed that full next-to-leading order analysis of
mesonic masses within the chiral perturbation theory does not
 constrain $\dm$ at all \quref{dgh}.  This is due to
the lack of knowledge of the electromagnetic contribution
to meson masses at next order in quark masses.
It is however not excluded that at this order
the Dashen Ansatz is violated.
Recent analysis of the decay $\eta \rightarrow 3 \pi$ suggests that
\beq
 (m^2_{{\rm K}^{\pm}}-m^2_{{\rm K}^{0}}-m^2_{{\pi}^{\pm}}
  + m^2_{{\pi}^{0}})_{\rm EM} = 1300 \pm 400 {\rm MeV}^2,
\eeq
in contrast to 0 for the Dashen Ansatz. If this were true
${\cal R}=0.34$ and $\dm = 4.2$~MeV.

Therefore we fix cutoff {\em vs.} $M$ dependence as in \qref{ab10},
where the details can found.
The experimental numbers which are used to fix the model
parameters are $f_{\pi}=93$~MeV, $m_{\pi}$ and $m_{\rm K}$. Then
it comes out that $m_{\rm u}+m_{\rm d}=12.2$ MeV,
$\ms\simeq 150$ MeV. The kaon decay constant is then constrained
to $f_{\rm K}=105$~MeV about 10\% below the experimental value.
The allowed ra ge for $\dm$ is then, as discussed above, 2.6 -- 4.2 MeV,
however the lower values are rather unlikely, since they come out
by completely ignoring the electromagnetic contributions.

\section{Solitons and the Quantization of Zero Modes}

The effective Euclidean action for the semibosonized SU(3) NJL model
reads (after integrating out the quark fields) \cite{bolo:ab4,bolo:wealre}:
\beq
S_{\rm eff}= -{\rm Sp~ ln}(-i \dsl + m + M U^{\gamma_5}). \label{eq:Seff}
\eeq
Here $M$ is the constituent quark mass, $m$ the current quark mass matrix
and $U^{\gamma_5}$ describes the 8 Goldstone modes of the SU(3) chiral
symmetry.
The real part of
Eq.(\ref{eq:Seff}) is assumed to be regularized with the proper-time
regularization function of \qref{ab4} and Sp denotes the
functional trace.

First we make use of the  trivial embedding of
\quref{wit83b} of the SU(2) chiral field  $U_0(x)=(\sigma_{(2)} +
i\gamma_5{\vec\tau}{\vec\pi})/f_\pi$ into the isospin subgroup of SU(3)
according to
\beq   U(x) = \left( \matrix{ &U_0 &0 \cr &0  &1\cr} \right) .
\label{g3}
\eeq
The
soliton solutions of SU(2) are also solutions for SU(3) and the
embedding (\ref{g3})
gives the correct constraint on the Hilbert space of the baryonic
states \quref{wit83b}.

The soliton solutions are found by employing the {\em hedgehog}
Ansatz for the field $U_0$. The details are widely described in the
literature.
Following the treatment of
\qref{anw} we quantize the soliton by introducing time dependent
rotations of the {\em hedgehog} field:
$ U(x,t)=A(t)\; U(x)\;A(t)^\dagger $.
 This rotation can be undone by
rotating the quark fields:
${\tilde q}=A(t) q$ and  ${\tilde{\bar q}}={\bar q}
A(t)^\dagger$. Defining "angular velocities" $\Omega$:
\bea
A^\dagger \dot{A} &=&
 i\Omega_E ~~=~~\frac{i}{2}\lambda_a \Omega_E^a
 \eea
one can rewrite $S_{\rm eff}$ as:
\beq
S_{\rm eff}= -\Sp ~{\rm ln}
(\partial_\tau+H+i\Omega_E-i\gamma_4A^\dagger mA). \label{eq:Seff1}
\eeq
The following relation between Euclidean and Minkowski velocities holds:
$i\Omega_E=\Omega_M$ and $\Omega_E^\dagger=\Omega_E$.

Expanding (\ref{eq:Seff1})     up to
the quadratic order in
$\Omega$ (in Minkowski metric and in the chiral limit) one gets \qref{ab4}:
\beq
L_0 =-M_{\rm cl}+\frac{1}{2} \Omega_{a}I_{ab}\Omega_{b} \label{eq:L}
\eeq
where tensor of inertia
$I_{ab}={\rm diag}(I_1,I_1,I_1,I_2,I_2,I_2,I_2,0)$
 can be found in \qref{ab4}.

To calculate the  mass splittings  one has to expand the effective
action in powers of the current quark mass $m=
\mu_{0}\,\lambda_{0} - \mu_{8}\,\lambda_{8} -\mu_{3}\,\lambda_{3}
$ ($\lambda_{0}=\sqrt{2/3} \;{\bf 1}$) with:
\beq
 \mu_{0}  =  \frac{1}{\sqrt{6}}(m_{\rm u}+m_{\rm d}+m_{\rm s}),
{}~~\mu_{8}  = \frac{1}{\sqrt{12}}(2\,m_{\rm s}-m_{\rm u}-m_{\rm d}),
{}~~\mu_{3}  =  \frac{1}{2}\Delta m
\eeq
where $\Delta m=m_{\rm d}-m_{\rm u}$.

Expanding the effective action up to
terms of the order of $\ms$, $\dm$,  $\ms^2$, $\dm\; \ms$,
$\ms\; \Omega$ and  $\dm\; \Omega$
one gets:
\begin{eqnarray}
L_{m}~ & = &-\sigma\; m_{\rm s} +
             \sigma \left(  m_{\rm s} D_{88}^{(8)}
	         + \frac{\sqrt{3}}{2}\; \dm D_{38}^{(8)}\right), \nonumber\\
L_{m \Omega} & = & -\frac{2}{\sqrt{3}} m_{\rm s}\; D_{8a}^{(8)}
K_{ab}{\Omega}_{b} - \dm D_{3a}^{(8)} K_{ab}{\Omega}_{b},
  \nonumber \\
L_{m^2} & = & \frac{2}{9} m_{\rm s}^2 \left(
N_{0} (1-D_{88}^{(8)})^2+ 3 N_{ab} D_{8a}^{(8)} D_{8b}^{(8)}\right)\nonumber\\
 & &+ \frac{2}{3\sqrt{3}} m_{\rm s}\;\dm \left(
N_{0} (D_{38}^{(8)} D_{88}^{(8)} - D_{38}^{(8)}   )
+ 3 N_{ab} D_{3a}^{(8)} D_{8b}^{(8)}\right),
\label{eq:LmOm}
\end{eqnarray}
where the constant $\sigma$
is related to the  nucleon sigma term
$\Sigma=3/2\;(m_{\rm u}+m_{\rm d})\;\sigma$ and
$D_{ab}^{(8)}=1/2\;{\rm Tr}(A^{\dagger}\lambda_a A \lambda_b)$.
The mass spectrum obtained with the help of $L_0+L_m+ L_{m \Omega}$
was discussed in
Refs.\cite{bolo:ab4,bolo:ab5}; there one can also find explicit
formulae for
$K_{ab}={\rm diag}(K_1,K_1,K_1,K_2,K_2,K_2,K_2,0)$.
Let us here only remind that the  {\it anomalous} moments of inertia $K_i$
are nearly entirely given by the valence part, whereas the contribution
of the valence level to $I_i$ amounts to approximately 60\%.
The quantities
$N_{ab}={\rm diag}(N_1,N_1,N_1,N_2,N_2,N_2,N_2,N_0/3)$
have been derived in \qref{ab10}.
Their values together with the values of $I_{1,2}$ and $K_{1,2}$
for different constituent masses are listed in
\qtab{tabIKN}.

The lagrangian of Eq.(\ref{eq:LmOm}) reminds the one of the Skyrmion.
The quantization proceeds as in the Skyrme model
and has been  described in detail in the literature \quref{pw2}.
Let us here remind that at first
 one defines the quantities:
\beq
J_a=I_{ab}\Omega_b-\mu_i
D_{ib}K_{ba}-\delta_{a8}\frac{\Nc}{2\sqrt{3}}
\label{eq:Ja}
\eeq
($i=3$~and~$8$, $a,b=1\ldots 8$)
which, as a result of the quantization procedure,
are promoted to the spin operators ${\hat J}_a$. Note
that the relation (\ref{eq:Ja}) depends on the quark masses.
The wave function of the baryon state $B=Y,T,T_3,J,J_3$
belonging to the SU(3) representation ${\cal R}$ reads (see Appendix A
of \qref{ab10}):
\begin{eqnarray}
\mid {\cal R},B >
 & = &\sqrt{{\rm dim}{\cal R}}
\left< Y,I,I_{3} \mid D^{({\cal R})}(A) \mid -Y^{\prime} ,J,-J_{3}
\right>^{\bf *},
\label{eq:wf}
\end{eqnarray}
where the right hypercharge $Y^{\prime}$ is in fact constrained to be
$ -1$.
The lowest SU(3) representations which contain
states with $Y=1$ are: ${\cal R}=${\bf 8} and ${\cal R}=${\bf 10}.
The quantized hamiltonian from \qeq{eq:L} reads:
\beq
H^{(0)} = M_{\rm cl} + H_{\rm SU(2)} +H_{\rm SU(3)} ,
\label{ham0}
\eeq
\begin{eqnarray}
H_{\rm SU(2)}=     \frac{1}{2I_{1}}
C_{2}( {\rm SU(2) })
, &~~~~~~~~ &
H_{\rm SU(3)}=
\frac{1}{2I_{2}}  \left[
C_{2} ({\rm SU(3)})-C_{2}({\rm SU(2)})-\frac{N_{\rm c}^{2}}{12}
\right]
.   \nonumber
\end{eqnarray}
Here $C_2$ denote the Casimir operators of the spin SU(2) and flavor
SU(3). $M_{\rm cl}$ is the classical soliton mass. It has been
calculated by many authors and its value turns out to be relatively large:
$M_{\rm cl}\approx 1.2$~GeV. This is a common problem for all chiral models.
 There are however some negative corrections to it, like Casimir energy
or rotational band corrections which
might bring $M_{\rm cl}$ to the right value.
In this paper, instead on
insisting on the calculation of the absolute masses, we will
concentrate on the mass splittings.
\newpage

\begin{table}
\caption{Moments of inertia for different constituent masses}
\label{tabIKN}
\begin{center}
\begin{tabular}{ccccccccr}
$M$&$\Sigma$~[SU(2)]&$I_1$ & $I_2$ & $K_1$ & $K_2$ &$N_0$&$N_1$&$N_2$
\\
{}~$$[MeV] & [MeV] & [fm] & [fm] & [fm] & [fm] & [fm] & [fm] & [fm]   \\
\hline
 363. & 60.32& 1.512& 0.720& 0.606& 0.372& 0.765& 0.647& 0.496  \\
 395. & 58.14& 1.285& 0.618& 0.438& 0.290& 0.704& 0.500& 0.408  \\
 419. & 56.14& 1.178& 0.569& 0.369& 0.255& 0.668& 0.438& 0.370  \\
 423. & 55.52& 1.156& 0.560& 0.357& 0.250& 0.658& 0.426& 0.362  \\
 442. & 52.52& 1.070& 0.521& 0.315& 0.229& 0.603& 0.379& 0.329  \\
\end{tabular}
\end{center}
\end{table}

The hamiltonian up to terms linear and
quadratic in $m_{\rm s}$ and linear in  $\Delta m$
and $\ms \Delta m$ reads:
\bea
H^{(1)}&  = & \left\{ \sigma
       - r_2 Y
       - (\sigma-r_2) D_{88} +
      \frac{2}{\sqrt{3}} (r_1-r_2)
      \sum\limits_{A=1}^3 D_{8A} \hat{J}_A  \right\} \ms, \nonumber \\
H^{(2)}_{\rm kin} &  = & \frac{2}{3} \left\{
         r_2 K_2  ( 1 - D_{88}^2 )
       + (r_1 K_1-r_2 K_2)
       \sum\limits_{A=1}^3 D_{8A}^2  \right\} \ms^2,   \nonumber \\
H^{(2)}_{\rm dyn} & = & -\frac{2}{9} \left\{ (N_0+3 N_2)
           -2 N_0 D_{88}
       +  (N_0- 3 N_2) D_{88}^2
       +  3 (N_1-N_2) \sum\limits_{A=1}^3 D_{8A}^2
       \right\} \ms^2, \nonumber \\
h^{(1)}&  = & \left\{
       - r_2 T_3
       - \frac{\sqrt{3}}{2}(\sigma-r_2) D_{38} +
      (r_1-r_2)
      \sum\limits_{A=1}^3 D_{8A} \hat{J}_A  \right\} \dm, \nonumber \\
h^{(2)}_{\rm dyn} & = & \frac{2}{3\sqrt{3}} \left\{ N_0 D_{38}
         (3 N_2-N_0) D_{38} D_{88}
       +  3 (N_2-N_1) \sum\limits_{A=1}^3 D_{38} D_{8A}
       \right\} \ms \dm, \nonumber \\
h^{(2)}_{\rm kin} &  = & \frac{2}{3} \left\{
       -  r_2 K_2   D_{38} D_{88}
       + (r_1 K_1-r_2 K_2)
       \sum\limits_{A=1}^3
        D_{38} D_{8A}  \right\} \ms \dm, \label{eq:hams}
\eea
where $r_i=K_i/I_i$ ($i=1,2$) and $T_3$ stands for isospin.
We have split the $O(\ms^2)$ and $O(\ms\dm)$
hamiltonian into
the {\it kinematical} part which appears due to the fact that the
quantization relation between the angular velocity and the spin operators
$\hat J$ is $\ms$ and $\dm$ dependent (see Eq.(\ref{eq:Ja}))
and the {\it dynamical} part
which comes from the expansion of the effective action in terms of $m$.

The hamiltonians $H^{(1)}$ and $h^{(1)}$ mix states of different
SU(3) representations.
The corresponding
$O(\ms^2)$ contribution to the energy reads:
\bea
E^{(2)}_{\rm wf}&=&- \left\{\frac{1}{60}
\left(1+ Y -X+ \frac{1}{2} Y^2 \right)
(\sigma-r_1)^2
\right.
\nonumber \\
 & &~~~\left.
 +\frac{1}{250} \left( \frac{13}{2}+\frac{5}{2}X -\frac{7}{4} Y^2 \right)
 \frac{1}{9}(3 \sigma + r_1 - 4 r_2)^2
 \right\} I_2 \ms^2 \label{eq:wf8}
 \eea
 for the octet and for the decuplet:
\bea
 E^{(2)}_{\rm wf} &=&- \left\{ \frac{1}{16}
 \left(1+ \frac{3}{4}Y+ \frac{1}{8} Y^2 \right)
 \frac{1}{9}(3 \sigma - 5 r_1 + 2 r_2)^2
 \right.
 \nonumber \\
 & &~~~\left.
 +\frac{5}{336} \left( 1 - \frac{1}{4} Y - \frac{1}{8} Y^2 \right)
(\sigma + r_1 -2 r_2)^2
\right\} I_2 \ms^2. \label{eq:wf10}
 \eea
Here
$X=1-T(T+1)+1/4\;Y^2$ is the usual combination entering
Gell-Mann--Okubo mass relations ($T$ stands for isospin).

Similarly one can write the $O(\ms \dm)$ wave function contribution
to the octet states:
\bea
e^{(2)}_{\rm wf} & = & \frac{1}{30}(\sigma-r_1)^2\;(1+Y)\;T_3\,
                    ~I_2 \ms \dm \nonumber \\
	       &   &-\frac{2}{125}(\sigma+\frac{1}{3}r_1
	       -\frac{4}{3} r_2)^2 \;Y\;T_3\, ~I_2 \ms \dm. \label{e2}
\eea

\section{Hyperon Splittings}

With the help of the  matrix elements of the $D$ functions and spin
operators
one arrives at the following result for the hyperon splittings:
\bea
\Delta M^{(8)} &=& A -\frac{F}{2}\;Y-\frac{D}{\sqrt{5}}\;X - G\;Y^2,
\nonumber \\
\Delta M^{(10)} &=& B -\frac{C}{2\sqrt{2}}\;Y- H\;Y^2.
\label{abcdfgh}
\eea
Constants $A$ and $B$ do not contribute to the splittings within the
multiplets, however they shift the mass of the centers and contribute to the
{\bf 10-8} mass difference. Constants $G$ and $H$, not present in the first
order Gell-Mann--Okubo mass formula, are of the order of $\ms^2$.

Hyperon splittings obtained with the help of Eq.(\ref{abcdfgh}) have
been discussed in detail in \qref{ab10}. Here for completeness we repeat
only the main points.

Experimentally one gets:
\bea
F &=&\Xi-{\rm N}=379~{\rm MeV},\nonumber \\
D & =&\frac{\sqrt{5}}{2}(\Sigma-\Lambda)=86~{\rm MeV},\nonumber \\
G&=&\frac{1}{4}(3\Lambda+\Sigma)-\frac{1}{2}({\rm N}+\Xi)=6.75~{\rm MeV}
\label{eq:8exp}
\eea
for the octet. For the decuplet the three operators: 1, $Y$ and $Y^2$ do
not form a complete basis and therefore there are two independent
relations which determine constants $C$ and $H$ with small uncertainty:
\bea
C=\sqrt{2}(\Xi^* - \Delta)&=&\frac{1}{\sqrt{2}}(\Omega-2 \Delta+\Sigma^*)=
422.5\pm3.5~{\rm MeV}, \nonumber \\
H=\frac{1}{2}(2\Sigma^*-\Xi^*-\Delta) & = &
\frac{1}{6}(3\Sigma^*-2\Delta-\Omega)=2.83\pm 0.33~{\rm MeV}.
\label{eq:10exp}
\eea

In \qtab{tabABCDFGH}
we list the coefficients $A\ldots H$ for a typical value of
$\ms=180$~MeV
as functions of the constituent mass $M$. It can be seen
that in order to reproduce the experimental numbers of
Eqs.(\ref{eq:8exp},\ref{eq:10exp}) one has to take
the constituent mass of the
order of 400~MeV. Then all constants $A\ldots H$ are roughly reproduced.
The constant $G$ and $H$ being of the order $O(\ms^2)$ are small.
For reasonable
strange quark  masses $O(\ms^2)$ corrections
to $A,B,C$ and $D$
are of the order of 20\%
of the leading $O(\ms)$ terms with the exception of $F$ for which
$O(\ms^2)$ corrections are almost zero. This is illustrated in
Fig.1 where $O(\ms^2)/O(\ms)$ contributions to constants
$A\ldots F$ are plotted as functions of $\ms$ for the fixed
value of $M=420$~MeV.
\begin{table}
\caption{Different contributions to coefficients of Eqs.(18)
for  $M=423$ MeV and $\ms=180$~MeV }
\label{tabABCDFGH}
\begin{center}
\begin{tabular}{cccccccc}
  & $O(\ms)$ &\multicolumn{4}{c}{$O(\ms^2)$} & total & exp. \\
  \cline{3-6}
  &        &    kin.  &   dyn.    &   w.f. & total  &         &    \\
  \hline
 A &  546.10 &   10.94 &  -64.64 &   -0.15 &  -53.85 &  492.25 &  -- \\
 B &  546.10 &   10.85 &  -64.55 &  -53.81 & -107.50 &  438.60 &  --  \\
 F &  381.20 &    1.18 &  -27.67 &   22.76 &   -3.73 &  377.47 &  379.00 \\
 D &  120.76 &   -0.02 &  -11.78 &   -0.07 &  -11.87 &  108.89 &   86.00 \\
 C &  348.16 &    1.20 &  -19.97 &   90.92 &   72.15 &  420.32 &  422.00 \\
 G &    0.00 &    0.61 &   -0.66 &    1.53 &    1.48 &    1.48 &    6.75 \\
 H &    0.00 &   -0.29 &    0.41 &    4.57 &    4.70 &    4.70 &    2.83 \\
\end{tabular}
\end{center}
\end{table}

\begin{figure}
\hspace{2 cm}\epsfbox{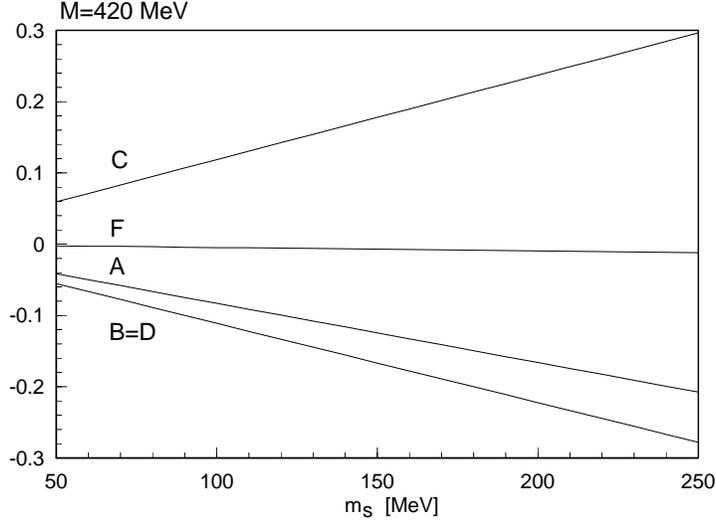}
\caption{Ratios of $O(\ms^2)/O(\ms)$ contributions to $A\ldots D,~F$
as functions of $\ms$}
\end{figure}

In order to make phenomenological statements
we adopt the following procedure: first for given
$M$ we find the optimal $\ms$  which reproduces {\bf 10-8} splitting.
To this end we define the mean octet and decuplet values:
${\overline M}^{(8)} = 1/2\;(\Lambda + \Sigma)=1155$~MeV and
${\overline M}^{(10)} = \Sigma^*=1385$~MeV. Then
$\Delta_{10-8}\equiv {\overline M}^{(10)} -{\overline M}^{(8)} =230$~MeV
is given by:
\beq
\Delta_{10-8}=\frac{3}{2 I_1} + B - A. \label{eq:10-8}
\eeq
Since $A-B={\rm const.}\times \ms^2$ one can numerically solve
Eq.(\ref{eq:10-8}) for $\ms$. The result is plotted in Fig.2
(see also \qtab{tabfd}).

\begin{figure}
\hspace{2 cm}\epsfbox{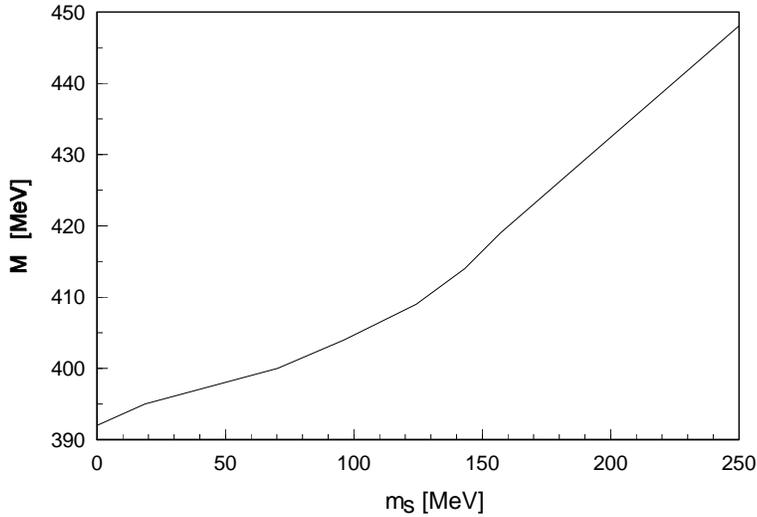}
\caption{$M$ {\em vs.} $\ms$ dependence induced by
the condition $\Delta_{10-8}=230$ MeV}
\end{figure}

\newpage

In Fig.3 we show the $\ms$ dependence of the deviations
{\it theory -- experiment} for each hyperon.
One should remember that for each $\ms$
the {\em optimal} constituent quark mass $M$ was used, so that
$\Delta_{10-8}$ was automatically reproduced for each $\ms$.
 The smallest deviations $\pm 7$~MeV for all splittings correspond to
$\ms\simeq 185$~MeV, {\it i.e.} $M\simeq~426$~MeV.

\begin{figure}
\hspace{2 cm}\epsfbox{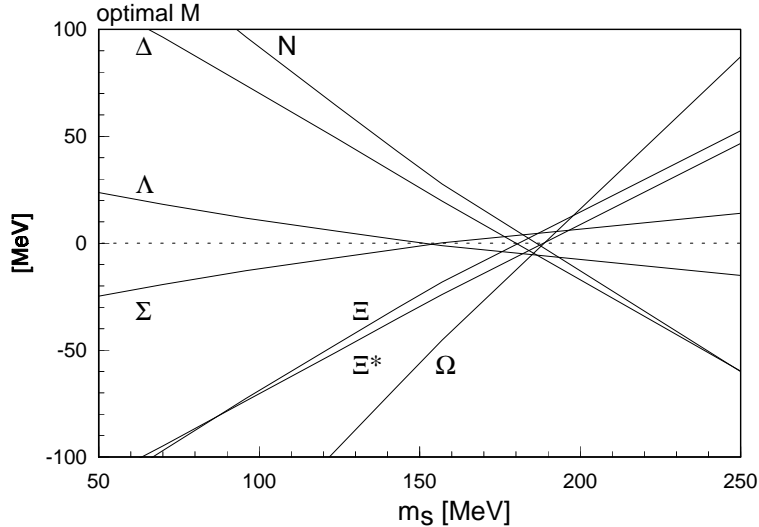}
\caption{The $\ms$ dependence of the deviations
{\it theory -- experiment} for each hyperon
for $M$ chosen according to Fig.2}
\end{figure}

It is constructive to plot the dependence of constants $F$, $D$ and
$C$ entering the Gell-Mann--Okubo mass relations as functions of
$\ms$ (and {\em optimal} $M$). Fig. 4 shows the influence of
the $O(\ms^2)$ terms on these quantities. Dashed lines represent
the first order results and  solid lines the full result. It
can be seen that $O(\ms^2)$ corrections substantially improve
model predictions. $F$ hits its experimental value for $\ms=181$ MeV
and the corresponding $M\simeq 424$ MeV. For these values $C$ is within
the experimental errors and $D$ overshoots slightly the experimental
value.

In this way we have fixed $M$ -- the only remaining free parameter of
the model to the value of about 425 MeV.
The corresponding value of $\ms \simeq 180$ MeV required to fit the
{\bf 10-8} splitting  overshoots the value deducted from the meson sector
$\ms\simeq 150$ MeV. However, in an exact treatment of the vacuum sector
and perturbation theory around this exact vacuum for the soliton sector,
the crucial quantity is $M_s-M_u$, the difference of the constituent
quark masses, instead of $m_s$  and this turns out to be $\simeq 180-200\MeV\gg
m_s$.
Within such a treatment the discrepancy between meson  and baryon
sectors disappears \quref{ab15}.
\begin{table}
\caption{Constants $f$ and $d$ as functions of
$M$ and $\ms$ chosen according to Fig.2}
\label{tabfd}
\begin{center}
\begin{tabular}{cccccccccccc}
 $M$ &$\ms$&$f^{(1)}$&$f^{(2)}_{\rm wf}$&$f^{(2)}_{\rm dyn}$&
 $f^{(2)}_{\rm kin}$&$\sum f$
 &$d^{(1)}$&$d^{(2)}_{\rm wf}$&$d^{(2)}_{\rm dyn}$&
 $d^{(2)}_{\rm kin}$&$\sum d$ \\
\hline
 395.& 19.&3.34 &-0.05&-0.01&-0.00&3.28&0.46&0.01&-0.00&-0.001&0.47 \\
 419.&157.&3.21 &-0.34&-0.12&-0.02&2.72&0.45&0.07&-0.02&-0.006&0.49 \\
 423.&177.&3.18 &-0.37&-0.14&-0.02&2.65&0.45&0.07&-0.02&-0.007&0.49 \\
 428.&209.&3.11 &-0.41&-0.16&-0.02&2.53&0.44&0.08&-0.03&-0.007&0.48 \\
 442.&258.&3.03 &-0.45&-0.19&-0.02&2.37&0.43&0.09&-0.03&-0.008&0.48 \\
 \end{tabular}
\end{center}
\end{table}

\begin{figure}
\hspace{2 cm}\epsfbox{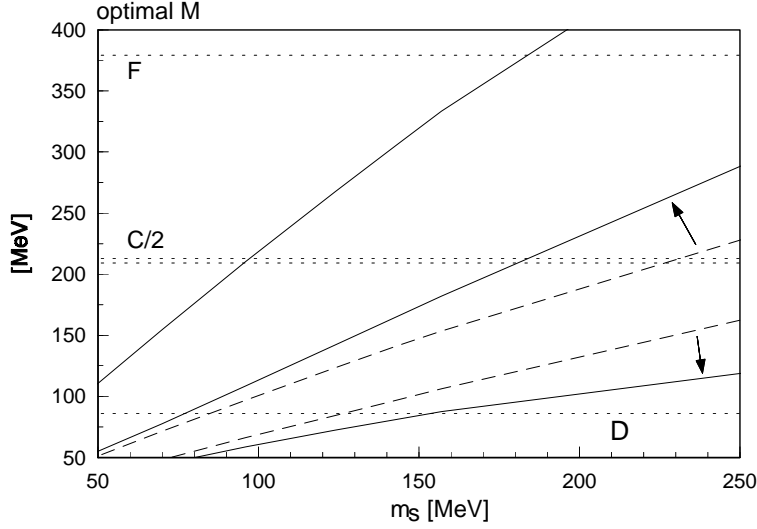}
\caption{$F$, $C$ and $D$ as functions of $\ms$; dashed lines -- first
order contributions, solid lines -- full $O(\ms^2)$ result, note almost
zero correction to $F$}
\end{figure}

\section{Isospin Splittings}

Hadronic parts of the isospin splittings have been estimated in
\quref{gale}:
\beq
(n-p)_{\rm H} = 2.05 \pm 0.3,~~
(\Sigma^- - \Sigma^+)_{\rm H} = 7.89 \pm 0.3,~~
(\Xi^- - \Xi^0)_{\rm H} =5.5 \pm 0.7  \label{expiso}
\eeq
(in MeV) for the octet states. In \qref{ab5}  we have also
estimated  the splittings for the decuplet on the basis of the
simple Dashen parametrization, which has proven to work equally well
for both decuplet and octet particles.
  For the purpose of this work, however, we will
concentrate entirely on the octet, since the numbers of Eq.(\ref{expiso})
are experimental, whereas the decuplet estimations, in view of the
lack of the data, were based upon
the theoretical guess, as mentioned above.

 From theoretical point of view the isospin splittings are described
by the formula analogous to the Gell-Mann--Okubo mass formula for
the hyperon splittings, namely \quref{ab5}:
\beq
(\Delta M)_{\rm H} = -\frac{1}{3}f\, T_3 + d \, Y\;T_3, \label{gmo}
\eeq
where, taking into account Eq.(\ref{expiso}), one gets:
\bea
f =11.33 \pm 1.14~{\rm MeV}, & ~~~~ & d = 1.73 \pm 0.38~{\rm MeV}.
\eea
On the other hand $f$ and $d$ can be directly evaluated in the
present model with the help of Eqs.(\ref{eq:hams},\ref{e2}):
\bea
f^{(1)} = \frac{3}{4} (\sigma + r_1 +2 r_2), &~~ &
f^{(2)}_{\rm kin} =\frac{1}{5} (2 r_2 K_2 - 3 r_1 K_1 )\; \ms, \nonumber \\
f^{(2)}_{\rm dyn} =-\frac{1}{5}(2 N_0 - 3 N_1 + 2 N_2)\;\ms,  & &
f^{(2)}_{\rm wf} =-\frac{1}{10} (\sigma - r_1)^2 \; I_2 \,\ms, \nonumber \\
d^{(1)} = \frac{3}{20} (\sigma - 3 r_1 +2 r_2), & &
d^{(2)}_{\rm kin} =\frac{1}{45} (2 r_2 K_2 - 5 r_1 K_1 )\; \ms,  \\
d^{(2)}_{\rm dyn} =-\frac{1}{45}(4 N_0 - 5 N_1 + 2 N_2)\;\ms,  & &
d^{(2)}_{\rm wf} = \left[\frac{1}{30} (\sigma - r_1)^2
- \frac{2}{125} \left(\sigma+\frac{r_1}{3}-\frac{4r_2}{3}\right)^2
\right]\; I_2\,\ms \nonumber
\eea
and
\bea
f=(f^{(1)}+f^{(2)}_{\rm kin}+f^{(2)}_{\rm dyn}+f^{(2)}_{\rm wf})\dm,
&~~&
d=(d^{(1)}+d^{(2)}_{\rm kin}+d^{(2)}_{\rm dyn}+d^{(2)}_{\rm wf})\dm.
\eea
It is worth noting that in the order $\ms \dm$ the Gell-Mann--Okubo
mass formula of Eq.({\ref{gmo}) is not violated, but the coefficients
$f$ and $d$ depend on $\ms$. In \qtab{tabfd}
we list the values of $f$ and $d$ for different constituent
masses $M$ and for $\ms$ chosen according to Fig 2.

It is clearly seen from \qtab{tabfd} that the second order corrections
to $f$ and $d$ are dominated by the wave function contribution, however
the dynamical part, although smaller,
 is by no means negligible. This is in contrast to
the hadronic splittings where the dynamical corrections were equally
important as the wave function ones \quref{ab10}.

Finally in Figs. 5-7 we show the $\dm$ dependence of the isospin
splittings for three constituent masses $M=419$, 423
and 428 MeV and three corresponding
$\ms=157$, 177 and 209 MeV respectively. The first set corresponds to the
strange quark mass as required by the meson sector. For this values,
however, the hyperon spectrum is not correctly reproduced (see Figs.
3 and 4). The second set corresponds to the best fit to the hyperon
spectra. The third set has been chosen to show what happens if $\ms$
starts overshooting the optimal value.
\begin{figure}
\hspace{2 cm}\epsfbox{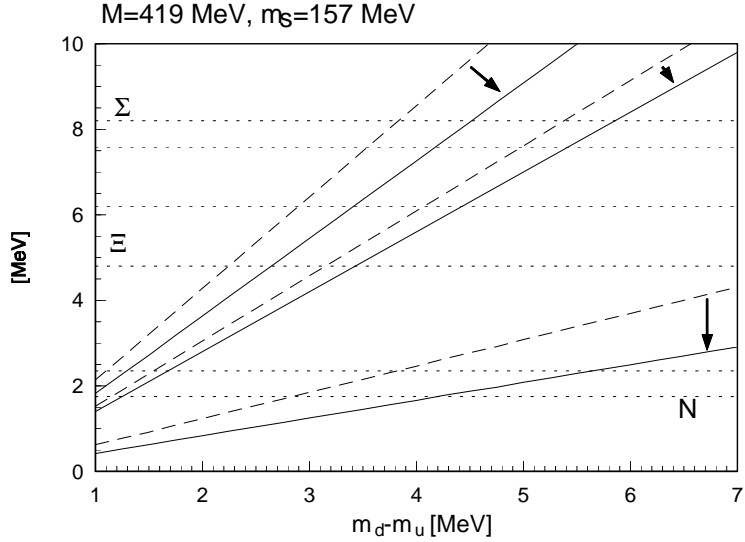}
\caption{$\dm$ dependence of the isospin splittings for $M=419$ MeV and
$\ms=157$ MeV. Horizontal dashed lines correspond to experimental error
bars}
\end{figure}
\begin{figure}
\hspace{2 cm}\epsfbox{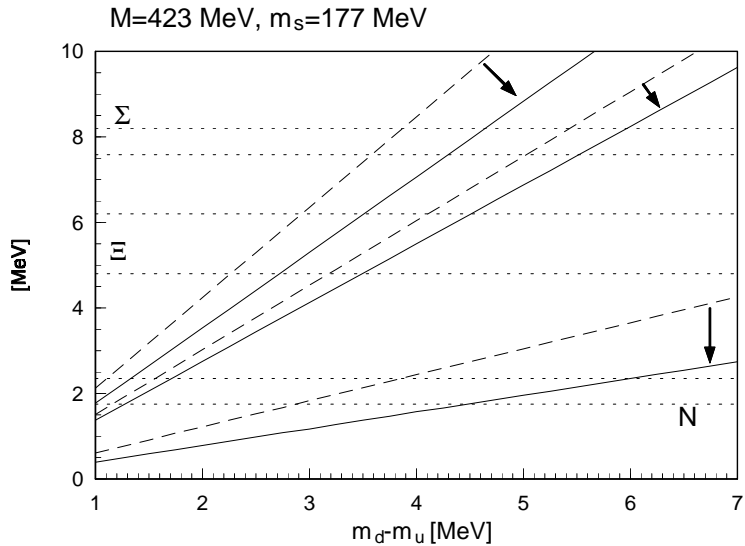}
\caption{Same as Fig.5 for $M=423$ MeV and $\ms=177$ MeV}
\end{figure}

\begin{figure}
\hspace{2 cm}\epsfbox{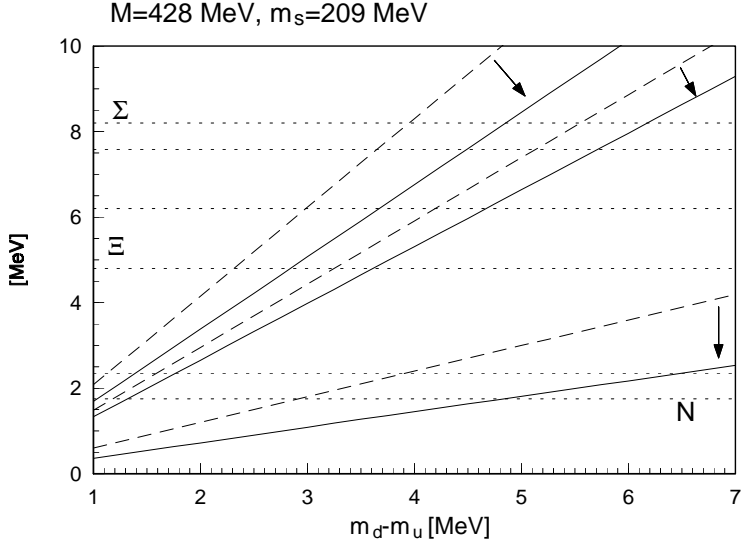}
\caption{Same as Fig.5 for $M=428$ MeV and $\ms=209$ MeV}
\end{figure}

The figures clearly show, that the slopes of the splittings
as functions of $\dm$ uniformly decrease with $\ms$. Without the
$O(\ms)$ corrections all splittings fall within the experimantal
error bars for $\dm \simeq 3.5$ MeV (dashed lines on Figs. 5-7).
When the $O(\ms)$ corrections are included the $\dm$ range for
which theoretical curves fall into experimental error bars
shifts towards higher values and shrinks at the same time,
so that for $M=428$ MeV (Fig. 7) there is no common value
of $\dm$ which would describe all splittings. For $M=423$ MeV
(Fig. 6) it is still possible to desribe all splittings wiith
$\dm\simeq 4.4$ MeV. This value should be compared with the
$\dm=2.6-4.2$ MeV which is needed to reproduce meson masses.

\section{Concluding Remarks}

Almost all chiral models have problems in reproducing isospin splittings
(see Ref.\C{bolo:Stern} and references therein). In our previous work
\C{bolo:ab5} we have
shown that in the semibosized NJL {\em all} isospin splitting
within the octet and decuplet of baryons are reproduced within
surprisingly good accuracy in the zeroth order in $\ms$. In the Skyrme
model the $O(\ms)$ corrections to the isospin splittings are large and
decrease the splittings. It has been therefore of importance to see how
this corrections influence the mass differences in the present model.
We have shown that the slope of the isospin
splittings as functions of $\dm$ decreses with increasing $\ms$. This
is in principle what happens also in the SU(3) Skyrme model, however here
the effect is much less pronounced. This is due to the existence of the
{\em anomalous} parts associated with the quantities $K_{1,2}$.
Already at the zeroth order, where $f^{(1)}=3.18$ for $M=423$~MeV the
{\em anomalous} contribution amounts to 28\%. The wave function
corrections  decrease $f$ as seen from \qtab{tabfd} by $-0.37$, however
without the {\em anomalous} part the decrease would be much stronger:
$-0.46$ for $\ms=177$~MeV. Altogether the final value of $f=2.65$ would
be decreased to $1.68$ ({\em i.e.} by 37\%) if the anomalous part was
absent. On the contrary, the influence of {\it anomalous} terms on the
coefficient $d$ is small; it amounts to 4\%.

Let us stress once more that the successful phenomenology of the
isospin splittings could not be achieved within the SU(2) NJL model,
where, similarly to the Skyrme model, the matrix elements of the
relevant operator simply vanish.

\vspace{1 cm}

\noindent{\bf Aknowledgements}

The work was partially supported  by {\it Alexander von
Humboldt Stiftung} and {\it Polish Research Grant} KBN~2.0091.91.01 (M.P.),
{\it Graduiertenstipendium des Landes NRW} (A.B.),
and the
{\it Bundesministerium fuer Forschung und Technologie} (BMFT, Bonn).
%

\end{document}